\documentclass[11pt,preprint]{aastex}

\def\avg#1{\langle #1 \rangle}

\def\msol{\hbox{$M_\odot$}}

\def\etal{{\it et al.~}}
\def\iso#1#2{\mbox{${}^{#2}{\rm #1}$}}
\def\be#1{\iso{Be}{#1}}
\def\li#1{\iso{Li}{#1}}
\def\b1#1{\iso{B}{1#1}}

\def\mej{M_{\rm ej}}
\def\ek{E_{\rm K}}

\def\ee#1#2{#1 \times 10^{#2}}

\def\rate#1{{\cal R}_{\rm #1}}
\def\snr{\rate{SN}}
\def\hnr{\rate{HN}}
\def\yield#1#2{\avg{m_{{\rm ej},#1}}_{\rm #2}}
\def\yldsn#1{\yield{#1}{SN}}
\def\yldhn#1{\yield{#1}{HN}}

\def\persn#1{\epsilon_{\rm #1}}

\def\pref#1{(\ref{#1})}

\def\beq{\begin{equation}}
\def\eeq{\end{equation}}
\def\beqar{\begin{eqnarray}}
\def\eeqar{\end{eqnarray}}

\begin{document}

\title{PRODUCTION OF LITHIUM, BERYLLIUM, AND BORON BY HYPERNOVAE}
 
\author{Brian D. Fields}
\affil{Center for Theoretical Astrophysics,
Department of Astronomy, University of Illinois,
Urbana, IL 61801, USA}

\author{Fr\'ed\'eric Daigne}
\affil{MPI f\"{u}r Astrophysik,
Karl-Schwarzschild-Str. 1, 85741 Garching bei M\"{u}nchen, Germany \\
also Institut d'Astrophysique de Paris, 98 bis Bd Arago 75014 Paris France}

\author{Michel Cass\'{e}}
\affil{Service d'Astrophysique, CEA, Orme des Merisiers,
91191 Gif sur Yvette, France \\
also Institut d'Astrophysique, 98 bis Boulevard Arago,
Paris 75014, France} 

\and 

\author{Elisabeth Vangioni-Flam}
\affil{Institut d'Astrophysique, 98 bis Boulevard Arago,
Paris 75014, France}

\begin{abstract}
We investigate a possible nucleosynthetic
signature of highly energetic 
explosions of C-O cores 
(``hypernovae,'' HNe) which 
might be associated with gamma-ray bursts (GRBs).
We note that 
the direct impact of 
C- and O-enriched hypernova ejecta 
on the ambient hydrogen and helium leads to
spallation reactions which can produce large amounts
of the light nuclides lithium, beryllium, and boron (LiBeB).  
Using analytic velocity spectra of the hypernova ejecta, 
we calculate the
LiBeB yields of different exploding C-O cores associated with
observed hypernovae.
The deduced yields are $\sim 10^3$ times higher
than those produced by similar (direct) means in normal Type II supernovae,
and are  higher than the commonly used ones arising from shock 
wave acceleration
induced by Type II supernova (SN) explosions. 
To avoid overproduction of these elements
in our Galaxy, hypernovae should be rare events, with 
$\la 10^{-3}$ hypernova
per supernova, assuming a constant HN/SN ratio over time. 
This rate is in good agreement with that of long duration
GRBs if we assume that the gamma-ray
emission is focussed with a beaming factor 
$\Omega/4\pi \la 10^{-2}$. 
This encouraging result supports the possible HN--GRB association.
Thus, Galactic LiBeB abundance measurements offer a promising
way to probe the HN rate
history and the possible HN-GRB correlation.
On the other hand, if
hypernovae are associated to very massive pregalactic stars
(Population III) they would produce a LiBeB pre-enrichment in
proto-galactic gas, which could show up as a plateau 
in the lowest metallicities of the Be-Fe relation
in halo stars. 
\end{abstract}


\section{Introduction}

An unusual class of very energetic supernovae 
(``hypernovae,'' hereafter HNe) has recently
been observed (Iwamoto et al.\ 1998, 2000).
Observationally, these events are identified
by their high luminosities and peculiar light curves.
Theoretically, these events seem to be the highly energetic
core collapse explosion of C-O cores 
(Iwamoto et al.\ 1998, 2000;
Woosley et al.\ 1999;
Nakamura, Mazzali, Nomoto, \& Iwamoto 2001;
Tan, Matzner, \& McKee 2001).
It has been suggested 
(Iwamoto et al.\ 1998;
Wheeler, Yi, H\"{o}flich, \& Wang 2000)
that these may be associated with 
at least some gamma-ray bursts (GRBs).  

The purpose of this paper is to point out that
these very energetic stellar explosions are good
sites for the copious production of the light elements lithium,
beryllium, and boron (LiBeB).
This occurs through the
collision of the HN ejecta with the circumstellar medium.
Fields et al.\ (1996) noted that such
nucleosynthesis occurs in the explosion of all supernova
(hereafter SN) ejecta,
when the fastest ejecta collide with the surrounding
medium and undergo spallation reactions.
Fields et al.\ found the LiBeB production is
particularly large for exploding C-O cores 
(resulting from, e.g., WR explosions or binary interactions).
Even so, for explosions of C-O cores with
``normal'' energies,
the net light element
yields are too small to significantly affect
the Galactic evolution of LiBeB.

As we will see, for hypernovae
the LiBeB production efficiency is much
higher, due to: (1) a surface composition of hypernovae
which is essentially
composed of C and O, ideal parent objects for spallation into lighter isotopes;
and (2)
a significant fraction of the outer envelope is propelled to high
velocities (energies higher than nuclear reaction threshold) due to
the very high kinetic energy released in their explosion.
For the case of hypernovae,
the astrophysical context is reasonably well-defined because there
are only two key 
physical parameters (explosion energy, ejected mass),
both of which are constrained by observations of
the supernova light curves. Adopting a calculated velocity (energy)
spectrum of the ejected C and O it is straightforward to evaluate the
absolute yield of light elements by spallation.  The only difficulty
is that the fast nuclei are slowed down in the course of their
propagation, and that the cross sections are energy dependent. The
procedure adopted to take into account these effect is explained in
Fields et al.\ (1996).

We combine our theoretical LiBeB yields 
with a simple model of Be chemical evolution
to quantify the hypernova contribution to
Galactic Be.
By comparing these results with observed Be abundance 
determinations in very metal poor stars in the halo of
our Galaxy, we place an upper limit on
the ratio of HNe to Type II SNe.
This limit holds
assuming a constant HN/SN ratio.
In addition, if we
assume a correlation between hypernovae and
gamma ray bursts, we can constrain the fraction of GRB that can be
identified as LiBeB producing hypernovae.

The term ``hypernova'' has been
used in different ways by different authors
(e.g., Paczy\'nski 1998;
Iwamoto et al.\ 1998, 2000), so it is important
to clarify the meaning used here.  
We define a hypernova as a 
core-collapse
explosion
whose detailed mechanism is unknown, whose kinetic energy is much
higher than usual, and whose envelope is dominated
by carbon and oxygen (rather than hydrogen and helium).
Such events are possibly associated with 
long-timescale GRBs, and we will discuss this possible
association in detail in \S \ref{sect:grb}.

\begin{table}[htb]
\caption{
Parameters of Hypernova Candidates
\label{tab:params}
}
\begin{tabular}{|l|cccc|l|}
\hline\hline
 & $M({\rm C-O})$  & $\mej$  & 
 $\ek$  & $v_* = (\ek/\mej)^{1/2}$  &  \\ 
Object  & $\msol$  & $\msol$ & $10^{51}$ erg  & $10^4$ km/s 
  & Reference \\
\hline
SN1994I$^{*}$  & 2.1  & 0.9  & 1  & 0.75  &  
   Nomoto et al.\ \cite{nomo1} \\
SN1994I($\times$10)  & 2.1  & 0.9  & 10  & 2.4  &
   \\
SN1994I($\times$30)  & 2.1  & 0.9  & 30  & 4.1 &
   \\
SN1997ef & 10  & 7.6  & 8  & 0.73  &
  Iwamoto et al.\ \cite{iwa00} \\
SN1998bw(a)  & 13.8  & 10.8  & 30  & 1.2  &
  Iwamoto et al.\ \cite{iwa98}  \\
SN1998bw(b)  & 6  & 4.6  & 22 & 1.5  & 
  Woosley et al.\ \cite{wes} \\
SNIa  & 1.4  & 1.4  & 1  & 0.60  & \\
\hline\hline
\end{tabular}\\
$^{*}$ The reference model of Fields et al. (1996).
\end{table}

\section{Production of LiBeB by Different Stellar Progenitors}

We now compute the production of LiBeB by hypernovae.
As seen in Table 1, the bulk properties of HNe are diverse.
In particular, the explosion energies and ejected masses
apparently span a considerable range.  
One would expect that the LiBeB yields are very sensitive to both of these
parameters.  We will show this to be the case,
and we will use analytic expressions to derive the 
scaling of the yields with these parameters.

Fields et al.\ (1996) noted that the fastest ejecta of a
supernova explosion have energies above the thresholds
for nuclear spallation reactions.  When these fast particles
interact
with the surrounding medium, they will therefore produce
LiBeB.  The light element production depends on the
velocity spectrum of the explosion, particularly that of 
the outermost layers.  
The LiBeB yields also scale as the
local ISM (target) density, 
while the irradiation timescale
is that of the ionization energy losses and thus
scale inversely with density.
These density effects cancel,
giving a LiBeB nucleosynthesis which is independent of the local
density but which does depend on the fast particle (and ISM)
composition.
The Fields et al.\ (1996) study is based on the numerical simulation of
the Type Ic event
SN 1994I (Nomoto et al.\ 1994), which is modeled as the
explosion of a 2.1 \msol\ C-O core.  
In this 
model the outermost and fasterst layers  (the only
ones that count in our problem) are a mixture of C and O
with no H and a small ($\sim 10\%$) admixture of He.

The velocity spectrum of the outermost layers can
be calculated analytically, as was shown by
Imshennik \& Nadyozhin (1988, 1989)
and reviewed in Nadyozhin (1994); 
these results have recently
been confirmed and extended into the relativistic
regime by Matzner \& McKee (1999).
Fields et al.\ (1996) give a full derivation of the relevant formulae for
our problem; here, we will only summarize key
inputs and results.
The particle spectrum is determined by the bulk hydrodynamics
of the problem, namely
the velocity (or kinetic energy) 
distribution as a function of mass shell
$M(>v)$.  
The ejected particles have a
energy spectrum 
$dN/d\varepsilon = m^{-2} \; v^{-1} \; dM/dv$
where $m$ is the mean particle mass.
Nadyozhin (1994) 
and Matzner \& McKee (1990) find that the velocity
spectrum of the fastest ejecta is a power law
\beq	
\label{eq:Mv}
M(>v) = \zeta^s \ \mej \left(v/v_*\right)^{-s}
\eeq
where $\mej$ is the ejected mass
and $v_* \equiv (E/\mej)^{1/2}$
is a characteristic speed associated with
the ejecta of an explosion having energy $E$.
The constants $\zeta$ and $s$ take on different
values depending on the polytropic index relevant
to the problem; for our case of 
$n=3$, we have $\zeta = 1.92$ 
and $s = 7.2$.
Eq.\ (\ref{eq:Mv}) holds for the fastest, outermost ejecta,
i.e., for $M(>v) \ll \mej$.

\begin{table}[htb]
\caption{
LiBeB Yields for Hypernova Candidates
\label{tab:yields}
}
\begin{tabular}{|l|ccccc|}
\hline\hline
 & \multicolumn{5}{|c|}{LiBeB Yield $\yldhn{i}$ (\msol)} \\
Object  & 
\li6  & \li7  & \be9  & \b10  & \b11 \\
\hline
SN1994I              & 0.13E-06 & 0.32E-06 & 0.40E-07 & 0.26E-06 & 0.12E-05\\
SN1994I($\times$10)  & 0.50E-03 & 0.13E-02 & 0.16E-03 & 0.10E-02 & 0.46E-02\\
SN1994I($\times$30)  & 0.26E-01 & 0.65E-01 & 0.83E-02 & 0.54E-01 & 0.24E+00\\
SN1997ef             & 0.88E-06 & 0.22E-05 & 0.28E-06 & 0.18E-05 & 0.81E-05\\
SN1998bw(a)          & 0.41E-04 & 0.10E-03 & 0.13E-04 & 0.84E-04 & 0.38E-03\\
SN1998bw(b)          & 0.12E-03 & 0.31E-03 & 0.39E-04 & 0.25E-03 & 0.11E-02\\
SNIa                 & 0.40E-07 & 0.10E-06 & 0.13E-07 & 0.82E-07 & 0.37E-06\\
\hline\hline
\end{tabular}
\end{table}

The key points here are that 
(1) the particles follow a steep power law spectrum
in kinetic energy per nucleon $\varepsilon$, with
$dN/d\varepsilon \propto \varepsilon^{-(s+1)/2}= \varepsilon^{-4.1}$
and (2) the fraction of ejected particles above
a particular velocity (or energy) threshold--and thus the fraction
available for LiBeB production spallation reactions--scales 
as the very strong power $v_*^s = v_*^{7.2}$.
Thus, once we adopt
the appropriate value of $s$,
the spectrum of particles is fixed, as are the
ratios among the LiBeB isotopes produced for
a given projectile and target composition.  These results are
(almost) independent of the explosion energy or ejected
mass, and thus should not vary much from one HN
to the next.\footnote{In fact, a dependence does remain
since the spectrum of eq.\ (\ref{eq:Mv}) is cut off
at an energy $E_{\rm max} \propto v_*^2$.
However, this is more difficult to calculate accurately
as it depends on the details of shock breakout.
Also, for the steeply falling spectra and high energies
we consider here, the results are only mildly
sensitive to $E_{\rm max}$.}
By contrast, the total LiBeB yield,
e.g.,  
$\yldhn{{\rm Be}}$,
depends
very strongly on the explosion energy and ejected mass,
with scaling 
\beq
\label{eq:yld_scale}
\yldhn{{\rm Be}} \propto \mej^{(s-2)/2} E^{s/2} = \mej^{-3.1} E^{3.6}
\eeq
and thus we can expect strong variations in LiBeB yields among
HNe.

So if the hypernova energy
is a factor of 10 higher than the usual $10^{51}$ erg, then
the mass ejected above LiBeB thresholds -- and thus the yields -- 
goes up by a factor of $10^{3.6}$ = 4000. 
This suggests that (low-mass) hypernovae can be prolific LiBeB sources.
Given the composition and spectrum of the projectiles and the known
composition of the target, one calculates the spallation yield in the
thick target approximation (excellent in our problem), taking into
account the energy dependent spallation cross sections.

A grid of models, comprising various kinds of exploding C-O cores
(observed and not) is presented in Table 2.  
One can verify that the yields obey the scalings
given in eq.\ (\ref{eq:yld_scale}).
The most 
copious LiBeB
producers are obviously those with lower mass and higher kinetic
energy.  Type Ia SNe
are a relatively interesting source due to their frequency, but they are less
productive than low mass hypernovae, since they eject comparable
masses but have 10 times less energy.  The high energy explosions of
massive hypernovae is overcompensated by their heaviness.  
Except for the energy, normal Type Ic SNe are events very similar to the SN
1998bw massive hypernova.
The
extremely large LiBeB production by 
low mass HNe, if they exist, makes them the
most efficient LiBeB-producing events known.  As such
they could have played a role in the evolution of light elements in
the early Galaxy, and possibly the intergalactic medium if
there were Pop III HNe.  Note that the calculated B/Be
ratio (Table 2), around 30, is consistent with the same
ratio observed in stars all along the metallicity scale
(Duncan et al. 1997, Primas et al. 2000, Cunha et
al. 2001). Moreover the isotopic ratios of lithium
($\li7/\li6\simeq 2.1$) and boron ($\b11/\b10\simeq 4.1$)
are in good agreement with these observations.

\section{LiBeB Abundance Constraints on Hypernova Rates}

We now turn to the contribution of HNe to the Galactic evolution of LiBeB.
From this point of view, it is important to note that
HN represent a {\em primary} LiBeB production
mechanism.  That is, due to their self-produced C-O cores,
the HNe ejecta are always enriched in C and O,
and thus the yields of Be are essentially independent of
the ambient interstellar medium metallicity.  
Consequently, we expect a linear scaling between the HN
ejecta of Be and O,  ${\rm Be \propto O}$, and
thus a constant Be/O ratio in the Galaxy.
Of course, HNe are not the only primary mechanism; 
another is LiBeB production via metal-rich particles
accelerated in superbubbles (Vangioni-Flam et al. 2000).  In addition,
standard Galactic cosmic rays, with a composition which
reflects the ISM metallicity, give a {\em secondary}
contribution which does depend on the interstellar
metallicity, and so scales as ${\rm Be \propto O^2}$.

The relative contribution of primary and secondary processes
to Be nucleosynthesis thus depends on the Be-O relation.
Unfortunately, O/H is difficult to measure in cool stars,
and controversy has arisen as two different O/H (and O/Fe)
trends have been claimed.  If O/Fe changes in Pop II 
(e.g., Israelian \etal\ \cite{igr,isr2001};
Boesgaard, King, Deliyannis, \& Vogt \cite{boesOFe};
Mishenina, Korotin, Klochkova, \& Panchuk \cite{mkkp})
Fields et al. (2000) showed
that both primary and secondary components are needed,
with primary dominating at $\mbox{[O/H]} \la -1.5$,
and secondary dominating above.  On the other hand, 
if O/Fe is constant in Pop II
(e.g., Carretta, Gratton, \& Sneden \cite{carretta};
Fulbright \& Kraft \cite{fk}), 
then a primary source of
LiBeB dominates until the roughly solar metallicities
(Vangioni-Flam et al.\ 1998).
Thus, {\em regardless} of the O/Fe behavior, there is a need for
primary Be at some level; the quantitative amount does depend
on the details of O data.  In what follows we will consider
the implications of both possibilities for O/Fe.

One can place LiBeB in full chemical evolution context 
(e.g., Vangioni-Flam et al.\ 2000;
Fields \& Olive 1999)
but a simplified approach, appropriate for Pop II, allows
one to focus on the physics of the HN contribution to Be.
In this approximation, we neglect the (small) astration
of Be, and thus the primary production of LiBeB species $i$ is described by
\beq
\label{eq:rate}
M_{\rm gas} \frac{{\rm d}}{{\rm d}t} X_i \simeq \yldhn{i} \hnr + \yldsn{i} \snr
\eeq
where $\yldhn{i}$ is the mean mass in $i$ created by one HN,
and $\yldsn{i}$ is the same quantity for one (superbubble) SN;
the rates of each event are given by $\hnr$ and $\snr$.

Since we are considering primary production, the yields
are independent of the initial ISM metallicity,
and in fact eq.\ \pref{eq:rate} applies not only to primary
LiBeB but also to metals such as O and Fe.
Thus we can write
\beq
\label{eq:BeO}
\beta \equiv \frac{X_{\rm Be}}{X_{\rm O}} = 
\frac{\persn{HN} \yldhn{{\rm Be}} + \yldsn{{\rm Be}}}
     {\yldsn{{\rm O}} + \persn{HN} \yldhn{{\rm O}}}
\eeq
where we have assumed a constant ratio 
\beq
\persn{HN} = \hnr/\snr \ \ 
\eeq
which we will refer to as the ``HN rate parameter.''

Given information about spallation and stellar yields,
eq.\ \pref{eq:BeO} allows us to relate the observed
$X_{\rm Be}/X_{\rm O} \simeq 16/9 \ {\rm Be/O}$
to the hypernova rate parameter $\persn{HN}$:
\beq
\label{eq:eps_full}
\persn{HN} = \frac{\beta \yldsn{{\rm O}} - \yldsn{{\rm Be}}}
                {\yldhn{{\rm Be}} - \beta \yldhn{{\rm O}}}
\eeq
Unfortunately, eq. \pref{eq:eps_full} as it stands
is difficult to evaluate due to the model-dependence of
the superbubble $\yldsn{{\rm Be}}$ and the unknown nature
of the HN oxygen yield $\yldhn{{\rm O}}$.  We can still
make progress, however, by setting an upper limit to
$\persn{HN}$, as follows.  First, we note that the
largest possible oxygen yield is when the ejecta is
pure oxygen: $\yldhn{{\rm O}} \le \yldhn{{\rm tot}}$.
We also note that the HN contribution is maximized
when we ignore the SN contribution.  It thus 
follows that we may limit the HN rate parameter to be
\beqar
\nonumber
\persn{HN} & \le & \frac{\beta \yldsn{{\rm O}} - \yldsn{{\rm Be}}}
                {\yldhn{{\rm Be}} - \beta \yldhn{{\rm tot}}} \\
         & \le & \frac{\beta \yldsn{{\rm O}}} 
                {\yldhn{{\rm Be}} - \beta \yldhn{{\rm tot}}}
\label{eq:eps_up}
\eeqar

With eq.\ \pref{eq:eps_up} in hand, we can now place limits
on the HN rate parameter.
We adopt the SN oxygen yield
$\yldsn{{\rm O}} = 2 \msol$,
which is insensitive to the choice of initial mass function.
For the total hypernova ejected mass we adopt 
the large and thus conservative value
$\yldhn{{\rm tot}} = 10 \msol$.
Finally, we must adopt a  Be yield for HNe.
As Table 2 illustrates, the wide range of HN masses
and energies implies a huge range in Be yields,
spanning orders of magnitude.
We will adopt 
$\yldhn{{\rm Be}} \simeq 10^{-5} \msol$,
the lower of the two values found for
SN 1998bw in Table 2.  The energy and ejected mass
dependence for this value 
are as in eq.\ (\ref{eq:yld_scale}).

With these parameters, we can evaluate eq.\
(\ref{eq:eps_up}) once we have made 
a choice of $\beta$. As noted above, this
depends on the oxygen data.
The weaker limit to $\persn{HN}$ comes from the constant
O/Fe case, in which Be is primary over all of Pop II.
In this case, we have $\beta \sim 3 \times 10^{-8}$, and thus
our fiducial numbers give
\beq
\label{eq:BeP}
\persn{HN} \le \ee{6}{-3}
\eeq
This evaluation is coherent with the upper limit which
can be derived from the beryllium
abundance in extremely metal-poor stars, as observed with
the VLT by Primas et al. (2000).
On the other hand, if O/Fe varies, then the
relevant Be/O ratio is that of the primary component,
which Fields et al. (2000) showed to be $\beta \sim \ee{8}{-10}$.
As this is smaller, we get a tighter limit:
\beq
\label{eq:BePS}
\persn{HN} \le \ee{1.6}{-4}
\eeq

There are various ways one can physically interpret 
a limit to $\persn{HN}$.  If one attributes the origin of a HN
to a mass effect, then $\persn{HN}$ is essentially the
fraction (by number) of massive stars which become a HN.
If we assume that stars above some lower mass limit
$m > m_{\rm HN}$ become a HN, then
for a Salpeter mass function (with massive stars 
in the range $10 \msol \le m \le 100 \msol$)
we derive $m_{\rm HN} > 91 \msol$ from eq.\ \pref{eq:BeP},
and $m_{\rm HN} > 99 \msol$ from eq.\ \pref{eq:BePS}.
These lie at the upper edge of the allowed range,
reflecting the smallness of the HN contribution. 
The HN origin could also be related to additional physical
parameters, such as binary interactions or rotation. 
In this case, the
limit to $\persn{HN}$ would reflect not only a mass effect
but also the fraction of
systems where such conditions are present.

The variation in HN energy and ejected mass will also
have an important effect on the limits quoted.
Had we adopted weaker explosions, or more massive
ejecta, these would lead to lower Be yields and
thus weaker limits on $\persn{HN}$.
One could even imagine turning the problem around:
given an independent measurement of
$\persn{HN}$, one could use these limits to infer
the mean $v_*$ for HNe.

\section{On the Possible Association Between GRBs and Hypernovae}
\label{sect:grb}

The Burst and Transient Source Experiment (BATSE) on board the Compton
Gamma-Ray Observatory (CGRO) detected more than 2500 gamma-ray bursts
(hereafter GRBs) from 1991 to 2000. The distribution of these bursts
over the sky is highly isotropic but the number of faint bursts is
notably smaller than the expected number if the distribution of bursts
was homogeneous in a Euclidean universe (Fishman and Meegan 1995 and
references therein). These two facts provide strong evidence that GRBs
occur at cosmological distance (Paczy\'nski 1991). This cosmological
origin is now firmly established for the long bursts 
($> 2\ \mathrm{s}$) 
thanks the discovery of their afterglows made possible by
the Beppo-SAX satellite. These late and fading counterparts are first
detected in the X-ray range, then in the optical and later in the
radio range. About twenty optical afterglows have been discovered
to date. The redshifts of most of these events have been measured, 
and range from $z=0.433$
(GRB 990712) to $z\simeq 4.5$ (GRB 000131). 
These values represent either a
direct measure of the redshift of the afterglow or in a few cases the
redshift of the host galaxy.

The two most popular models for the source of GRBs associate them with
the coalescence of two compact objects (NS-NS or NS-BH, Eichler et
al.\ 1989; Paczy\'nski 1991; Narayan et al.\ 1992; Mochkovitch et
al.\ 1993) or the collapse of a very massive star into a black hole
(collapsar, Woosley 1993). Such collapsars could be either the
collapse of a single Wolf-Rayet star endowed with rotation, or the
merger of the core of a massive star with a black hole or a neutron
star, and should lead to hypernovae as defined in this paper.  The
recent observations of the optical
afterglows of long bursts and the observations of their host galaxies
provide several pieces of evidence 
in favor of the association with massive stars:
the indication of dust extinction in optical afterglows and gas
absorption in X-ray afterglows suggest that GRBs occur near
star-forming regions (Paczy\'nski 1998; Bloom et al.\ 1998). 
The first
direct evidence for the GRB-massive star association
comes with the supernova SN 1998bw which is probably
associated with GRB 980425 (Galama et al.\ 1998). 
GRB 980326 and GRB
970228 might also be associated with supernovae (Bloom et al.\ 1999;
Reichart 1999).

As the sample of long bursts with a determined redshift is still
small, the distribution of GRBs as a function of redshift
$z$ must be estimated
indirectly. This has been done by many authors (for a review, see
Piran 1999) who fit the observed peak flux distribution assuming a
given rate of bursts $\rho_{\mathrm{GRB}}(z)$. The other parameters
for such a calculation are the luminosity distribution of bursts
$\Phi(L)$, which is usually taken to be independent of $z$, the
assumed spectral shape for the GRBs and the usual cosmological
parameters. The burst rate obtained by this method is very
uncertain. Whereas the results of these calculations are weakly
sensitive to the adopted values of the cosmological parameters (Cohen
\& Piran 1995), it has been shown that the BATSE sample is not
large enough to distinguish between to extreme assumptions : a constant rate
\begin{equation}
\rho_{\mathrm{GRB}}(z)=\rho_{0}
\end{equation}
or a rate proportional to the cosmic star formation rate 
\begin{equation}
\rho_{\mathrm{GRB}}(z) \propto \rho_{\mathrm{SFR}}\ .
\end{equation}
This is true in particular when relaxing the assumption that GRBs are
standard candles (Krumholz et al.\ 1998), which was usually made for
the first calculations (Sahu et al.\ 1997; Wijers et al.\ 1998) but is
not supported by the observations.

In this paper we will only consider the case of long GRBs. As we are
interested in putting constraints on the association between GRBs and
hypernovae, we will assume that the burst rate is indeed proportional
to the cosmic star formation rate. We will use the results obtained by
Porciani \& Madau (2001). They did a recent estimation of the GRB rate
under this assumption. They used $\Omega_{\mathrm{M}}=0.3$,
$\Omega_{\mathrm{\Lambda}}=0.7$ and $H_{0}=65 h_{65}\
\mathrm{km/s/Mpc}$. The spectral shape of the GRBs was given by the so
called GRB-function (Band et al.\ 1993). This is a phenomenological
$4$-parameter function which is known to fit very well the observed
spectra. They used the following parameters : $\alpha=-1.0$ for the
low energy slope, $\beta=-2.25$ for the high energy slope and
$E_{\mathrm{b}}=511\ \mathrm{keV}$ for the break energy, which
corresponds to the typical values obtained by Preece et al.\ (2000) who
did a detailed spectral study of a large sample of long GRBs.  We know
from the few GRBs with a measured redshift that the luminosity of GRBs
is strongly variable but the luminosity distribution is very poorly
constrained. Porciani and Madau used a power-law distribution :
\begin{equation}
\Phi(L)= C \left(\frac{L}{L_{0}}\right)^{\gamma}
\end{equation}
where $C$ is correctly normalized to have
$\int_{0}^{+\infty}\Phi(L)dL=1$. With all these assumptions and using
different estimations of the SFR, they found that their best fits give
a GRB rate of about 
1-2 bursts per million Type II supernovae, i.e.,
\beq
\label{eq:grb4pi}
\persn{\rm GRB,4\pi} = \ee{1-2}{-6}
\eeq
Clearly, this is much lower than the HN parameter found in the
previous section.

However, this does not demand that 
we reject the possible association between HNe and GRB.
If the GRB emission is focussed in an opening angle $\Omega$, the
rate parameter in eq.\ (\ref{eq:grb4pi})
has to be corrected upward by a factor $(\Omega/4\pi)^{-1}$.  
To reconcile the rates, we require that
$\Omega/4\pi\simeq 2\,10^{-4}$--$10^{-2}$.
This range has a considerable overlap with observed broad
distribution of beaming factor which spans $\Omega/4\pi \sim
10^{-3} - 10^{-1}$ (Frail et al. 2001). 
Moreover, our estimation
of the hypernovae rate $\persn{\rm HN}$ scales 
with the HN parameters in the same way as the Be yield,
i.e., as $\mej (E/\mej)^{s/2}=\mej (E/\mej)^{3.6}$. 
This strong dependence means that a modest change in
$E/\mej$ produces a large shift in $\Omega/4\pi$.
Thus one might hope to turn the problem around,
and use an accurate measure of the mean $\Omega/4\pi$
to infer the mean $E/\mej$.

Despite this encouraging result, important uncertainties
remain, especially due to our poor knowledge of the explosion
mechanism.  
If only a subclass of GRB progenitors leads to a
hypernova as we have defined, then $\persn{\rm HN}$ has to be
compared to a fraction only of $\persn{\rm GRB}$ and the
constrain on the beaming angle or the $E/\mej$ ratio becomes more severe.
On the other hand, the opposite situation
cannot be excluded: the isotropic envelope expansion
that we associate with a HN is always present, but the GRB
is produced only when certain unknown conditions allow the
acceleration of an ultrarelativistic outflow. In this case
$\persn{\rm HN}$ has now to be compared with $\persn{\rm
GRB}$ divided by the fraction of explosions producing a
GRB and large beaming angle are allowed, even with the $E/\mej$
ratio that we have adopted here.

\section{Conclusion}

Motivated by recent theoretical and observational
interest in hypernovae, we have
considered the LiBeB production by these objects.
We find that the LiBeB yields are very sensitive
to the explosion energy and ejected mass.
If these parameters typically take values
as found for SN 1998bw, then the Be yields
can be very large due to the high explosion energy.

Using the yields found for SN 1998bw,
we have calculated the impact of HNe on
LiBeB evolution in the galaxy.  
HNe represent a primary source of Be, and
thus are constrainted by the observed
primary component of Be vs O.
Using the observed Be data at low metallicities,
we are thus able to place limits on
the HN/SN ratio.
If we further associate HNe with GRBs,
we can infer a limit on the beaming angle of the GRB
emission $\Omega/4\pi \la 10^{-2}$ which is consistent with
independent estimates.
This agreement is encouraging, though of course
significant uncertainties remain.

Under the simple assumption of a constant HN/SN ratio
which has been made to derive these limits,
there are potentially important consequences for
LiBeB evolution.  The Li-Fe relation shows a small
slope (Ryan, Norris, \& Beers 1999) which
is consistent  with standard GCR production 
of Li (Ryan et al.\ 2000) but within errors also allows room for
other primary Li contributions.   
As the Li-Fe relation is measured more precisely, one may be able
to detect or limit the Li production by HNe, in addition to
that of standard and superbubble cosmic rays.

The assumption that the HN/SN ratio is
constant would be true, if both arise from massive star
formation and a constant, universal initial mass
function. While this is probably the simplest assumption,
other scenarios are possible.
For example, if a first generation of HN associated 
with very massive stars (Population
III) has existed, it could have produced proto-galactic beryllium,
along with C and O.
If these Pop III HNe produce strictly C and O but 
little or no iron, then this Be component
would manifest itself under as a plateau in the Be-Fe
correlation at the lowest metallicity (but a linear, primary 
trend in Be-O).
Consequently, if this Be-Fe plateau were observed, 
it would not necessarily imply that BBN has contributed.
In this scenario, the HN rate would not follow the SN rate
during the Pop III phase.
Thus, this picture could be tested 
by measuring the cosmic SN and HN rates at high redshifts.

\acknowledgements
We warmly thank Robert Mochkovitch for illuminating
discussions. This work has been supported in part by PICS
1076 from the CNRS.

\end{document}